\documentclass{article} 
\usepackage[dvips]{graphicx}
\usepackage{amscd}
\usepackage{amsmath}
\usepackage{amsfonts}
\usepackage{amssymb}
\numberwithin{equation}{section}
\begin{document}
\title{A DIAGRAM FOR BIANCHI A-TYPES\\ {\small by} }
\author{E.L. Sch\"ucking,\; E. Surowitz,\; and \; J. Zhao}
\date{}
\maketitle
\begin{abstract}
A diagram for Bianchi spaces with vanishing vector of structure constants
(type A in the Ellis-MacCallum classification)
illustrates the relations among their different types
under similarity transformations.
The Ricci coefficients and the Ricci tensor are
related by a Cremona transformation.
\end{abstract}

\section{Introduction}
Luigi Bianchi studied the geometry of $3$-dimensional Riemannian manifolds admitting isometries of simply transitive Lie groups. He classified these spaces based on a classification of the structure constants for the Lie algebras of their isometries \cite
{Bian98}.

The tensor of the structure constants $c_{ijk}$
\begin{equation}
c_{ijk}=-c_{ikj}\, , \qquad i,j,k=1,2,3
\end{equation}
of the  real $3$-dimensional Lie algebras can be split into a vector $c_k$
\begin{equation}
c_k \equiv c_{iik}
\end{equation}
and a vector-free part $n_{i\ell}$
\begin{equation}\label{n}
n_{i\ell}=\frac{1}{2}\left[\,c_{ijk}-\frac{1}{2}\left(\delta_{ij}\,c_k-\delta_{ik}\,c_j\right)\,\right] \epsilon_{jk\ell}
\end{equation}
which is a rank two tensor density. It is easy to see that $n_{i\ell}$ is symmetric since
\begin{equation}
n_{i\ell}\,\epsilon_{ip\ell}=0\,.
\end{equation}
Here $\epsilon_{ijk}$ is the completely skew-symmetric Levi-Civit\'a symbol with
\begin{equation}
\epsilon_{123}=+1
\end{equation}
and we do not distinguish between contra- and covariant indices because we classify with the orthogonal group in the Lie algebra.

We want to show in the following that the vector-free Bianchi types
exhibit a simple relation between group and curvature properties
that can be exhibited in a simple diagram.
For diagrams of the Bianchi types see also \cite{Schmidt}, \cite{Rainer}.

\section{Homogeneous Spaces}
The metric of a homogeneous space is given by
\begin{equation}\label{metric}
ds^2=\omega_i\,\omega_i \, , \qquad i=1,2,\dots,n
\end{equation}
where the $\omega_i$ are $n$ left-invariant differential one-forms. Assuming vanishing torsion \'Elie Cartan's first structural equations are
\begin{equation}\label{first}
d\omega_i=\omega_{ij}\wedge\omega_j \, , \qquad \omega_{ij}=-\omega_{ji}\,.
\end{equation}
The $n(n-1)/2\;\,\omega_{ij}$ are known as the connection forms. They can be developed in terms of the solder forms $\omega_k$ giving
\begin{equation}\label{Ricci}
\omega_{ij}=g_{ijk}\,\omega_k \, , \qquad g_{ijk}=-g_{jik}\,.
\end{equation}
The $g_{ijk}$ were introduced by Giorgio Ricci Curbastro and are named after him the Ricci coefficients.

The Maurer-Cartan equations for the left-invariant differential forms $\omega_i$ are given by
\begin{equation}\label{MauCar}
d\omega_i=-\frac{1}{2}\,c_{ijk}\,\omega_j\wedge\omega_k\,,
\end{equation}
where $c_{ijk}$ are the structure constants of the simply transitive group of isometries. We have from \eqref{first} and \eqref{Ricci}
\begin{equation}
d\omega_i=-g_{ijk}\,\omega_j\wedge\omega_k
\end{equation}
and obtain now, by comparison with \eqref{MauCar}
\begin{equation}\label{relation}
c_{ijk}=g_{ijk}-g_{ikj}\,.
\end{equation}
This means that the structure constants are obtained as twice the skew-symmetric part of the Ricci coefficients with respect to their last two indices.

Writing the last equation $3$ times cyclically we obtain
\begin{equation}
c_{ijk}+c_{jki}-c_{kij}=2\,g_{ijk}\,.
\end{equation}
Since the structure constants are constant, i.e., position-independent, it follows that the Ricci coefficients for the Bianchi spaces are also constant. As we shall see, it is the Ricci coefficients which are the parameters of choice.

The Jacobi identities for the structure constants $c_{ijk}$ read
\begin{equation}\label{Jacobiidentity}
c_{\ell im}c_{m jk}+c_{\ell jm}c_{m ki}+c_{\ell km}c_{m ij} = 0\,.
\end{equation}
Contracting the indices $\ell$ and $i$ we get the identities \cite{SS}
\begin{equation}\label{Jacobicontraction}
c_i c_{ijk} = 0 \, .
\end{equation}

\section{The Ricci Tensor}
The curvature $2$-form $\Omega_{ik}$ is defined by
\begin{equation}\label{Riccicurvature}
\Omega_{ik}=d\omega_{ik}-\omega_{ij}\wedge\omega_{jk} \, , \qquad \Omega_{ik}=-\Omega_{ki}\,.
\end{equation}
The identities
\begin{equation}\label{Bianchi}
0=\Omega_{ik}\wedge\omega_k
\end{equation}
give then the equivalent of the Jacobi identities for the structure constants of the Lie algebra.

The orthonormal components of the Riemann tensor are given by
\begin{equation}
\Omega_{ik}=\frac{1}{2}\,R_{ik\ell m}\,\omega_{\ell}\wedge\omega_m\,.
\end{equation}
In terms of the constant Ricci coefficients we obtain from \eqref{Ricci} and \eqref{Riccicurvature}
\begin{equation}
\Omega_{ik}=\left(g_{ikj}\,g_{jm\ell}-g_{ij\ell}\,g_{jkm}\right)\,\omega_{\ell}\wedge\omega_m\,.
\end{equation}
We have thus
\begin{equation}
R_{ik\ell m}=g_{ikj}\left(g_{jm\ell}-g_{j\ell m}\right)-g_{ij\ell}\,g_{jkm}+g_{ijm}\,g_{jk\ell}\,.
\end{equation}
This gives for the Ricci tensor
\begin{equation}
R_{km}=R_{ikim}=g_{ikj}\left(g_{jmi}-g_{jim}\right)-g_{iji}\,g_{jkm}+g_{ijm}\,g_{jki}\,.
\end{equation}
The second term cancels the fourth term and we obtain
\begin{equation}\label{Riccitensor}
R_{km}=g_{kij}\,g_{mji}+g_{jii}\,g_{jkm}\,.
\end{equation}
We decompose now the Ricci coefficients in the same way as we did for the structure constants. We define the vector $g_i$ by
\begin{equation}
g_i=g_{ijj}
\end{equation}
and obtain from \eqref{relation}
\begin{equation}\label{vector}
c_k=g_k\,.
\end{equation}
We notice that the expression for the Ricci tensor becomes particularly simple if the vectors $c_k$ and $g_k$ vanish.

We specialize now to the $3$-dimensional case of the Bianchi spaces. We call the vector-free part of the Ricci coefficients
\begin{equation}\label{symmetric}
\gamma_{k\ell}=\frac{1}{2}\left[\,g_{ijk}+\frac{1}{2}\left(g_j\,\delta_{ik}-g_i\,\delta_{jk}\right)\,\right] \epsilon_{ij\ell}\,.
\end{equation}
This tensor density is also symmetric since
\begin{equation}
\gamma_{k\ell}\,\epsilon_{kp\ell}=0\,.
\end{equation}
Inverting \eqref{symmetric} we obtain
\begin{equation}\label{Riccicoefficients}
g_{pqk}=\gamma_{k\ell}\epsilon_{pq\ell}-\frac{1}{2}\left(g_q\,\delta_{pk}-g_p\,\delta_{qk}\right)\,.
\end{equation}
Inserting \eqref{Riccicoefficients} into \eqref{Riccitensor} we obtain for the Ricci tensor
\begin{equation}\label{Riccitensorcomplete}
R_{km}=\epsilon_{ki\ell}\epsilon_{mjp}\gamma_{\ell j}\gamma_{pi}+\epsilon_{jkp}\gamma_{pm}g_j+\frac{1}{2}\,\delta_{km}g_jg_j\,.
\end{equation}
The symmetric tensor densities $\gamma_{i\ell}$ and $n_{i\ell}$ are related by \cite{GOLDEN}
\begin{equation}\label{Riccistructure}
\gamma_{i\ell}=\frac{1}{2}n_{jj}\delta_{i\ell}-n_{i\ell}\,,\qquad n_{i\ell}=\delta_{i\ell}\gamma_{rr}-\gamma_{i\ell}\,.
\end{equation}
From \eqref{Jacobicontraction} and \eqref{n} we obtain
\begin{equation}
c_i n_{i\ell}=0\,.
\end{equation}
Inserting \eqref{vector} and \eqref{Riccistructure} the above identities become
\begin{equation}
g_i\,(\delta_{i\ell}\gamma_{rr}-\gamma_{i\ell})=0\,.
\end{equation}

\section{The Vector-free Case}
In the following we shall specialize to the vector-free case
\begin{equation}\label{vanish}
c_i=g_i=0\,.
\end{equation}
George Ellis and Malcolm MacCallum referred to these $3$-dimensional real Lie algebras as ``Class A'' \cite{EM}. MacCallum calls them also unimodular algebras \cite{M}. In this case the relation between the matrix of the Ricci coefficients and the matrix
of the Ricci tensor becomes particularly simple. Specializing \eqref{Riccitensorcomplete} with \eqref{vanish} we obtain
\begin{equation}\label{Riccivectorfree}
R_{km}=-\epsilon_{ki\ell}\epsilon_{mpj}\gamma_{ip}\gamma_{\ell j}\,.
\end{equation}
Introducing the matrices
\begin{equation}
\Gamma=\left\|\gamma_{ij}\right\|\,,\qquad \rho=\left\| R_{ij} \right\|
\end{equation}
we have for $\text{det}\left\|\gamma_{ij}\right\|\ne 0$
\begin{equation}
\rho=-2\,\Gamma^{-1}\,\text{det}\,\Gamma\,.
\end{equation}
Using orthogonal transformations to diagonalize the matrix $\Gamma$ we have with
\begin{equation}\label{Gamma}
\Gamma=\begin{bmatrix} \gamma_1 & 0 & 0 \\
                       0 & \gamma_2 & 0 \\
                       0 & 0 & \gamma_3
       \end{bmatrix}
\end{equation}
for the Ricci tensor
\begin{equation}\label{rho}
\rho=-2\,\begin{bmatrix}
 \gamma_2\gamma_3 & 0 & 0 \\
 0 & \gamma_1\gamma_3 & 0 \\
 0 & 0 & \gamma_1\gamma_2
         \end{bmatrix}\,.
\end{equation}
The $\gamma_j$ are real numbers. Under similarity transformations of the differential forms $\omega_j$ (consisting of rotations, reflections and multiplications with a positive number) the matrix $\Gamma$ of the Ricci coefficients can be classified
according to the relative signs of its diagonal elements.
We have the following $6$ possibilities
\begin{align}\label{class}
\text{A}&:\quad(+,+,+)\;\text{or}\;(-,-,-) \nonumber\\
\text{B}&:\quad(+,+,-)\;\text{or}\;(-,-,+) \quad\text{and permutations}\nonumber\\
\text{C}&:\quad(+,+,\,0\,)\;\text{or}\;(-,-,\,0\,) \quad\qquad \raisebox{-.8ex}{$''\qquad ''$}\nonumber\\
\text{D}&:\quad(+,-,\,0\,)\;\text{or}\;(-,+,\,0\,) \quad\qquad \raisebox{-.8ex}{$''\qquad ''$}\\
\text{E}&:\quad(+,\,0\,,\,0\,)\;\text{or}\;(-,0\,,\,0\,) \quad\qquad \raisebox{-.8ex}{$''\qquad ''$}\nonumber\\
\text{F}&:\quad(\,0\,,\,0\,,\,0\,) \qquad \qquad \qquad\qquad\qquad\qquad\qquad\qquad .\nonumber
\end{align}
Since $\Gamma$ is a tensor density and not a tensor the sign of its signature is not invariant under the similarity transformations of the metric \eqref{metric}.

This classification gives also a classification of the Bianchi spaces since the eigenvalues of the Ricci tensor are invariants. The cases A through E are distinct. It is remarkable that the combinations $\text{A}' \equiv (-,-,-)$,\; $\text{B}' \equiv \;(+
,+,-)$ and its permutations, and the possibilities C and D cannot occur for the Ricci tensor. Cases F and E describe both the flat Euclidean $3$-dimensional space. We show the classifications in Figure 1.

\begin{figure}[htb]\label{stereo}
\begin{center}
\includegraphics[width=0.80\textwidth]{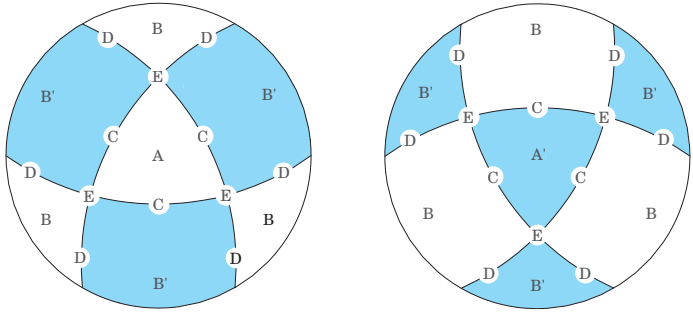}
\caption{The graph shows the stereographic projections of the northern and the southern
hemispheres from the south and the north poles of a sphere, respectively.
The sphere is divided by three great circles which are orthogonal to each other.
The different octants correspond to cases of different signs according to \eqref{class} while
the shaded octants together with their boundaries do not occur for the Ricci tensor.
A projective plane can be visualized by gluing together the two discs with rotation of
one by $\pi$.
Thus we get a disk in uniform shade, either white or grey, with divisions.
Topologically, \eqref{Riccivectorfree} can be viewed as a transformation from a sphere to the interiors of the divisions of a disk.
The boundary of the
disc corresponds to $\gamma_1+\gamma_2+\gamma_3=0$ which is invariant under
similarity transformations.}
\end{center}
\end{figure}
The vector-free Bianchi spaces and their Ricci coefficients depend on two real parameters under similarity transformations of the differential forms $\omega_j$. This is plausible since the matrix $\Gamma$ has six parameters and the similarity group has fo
ur. We thought it would be nice if one had a plane diagram that shows the relation of the different types of the Ricci coefficients and Ricci tensors. Instead of a diagram for the matrix $\Gamma$ one could consider one for the homogeneous quadratic polynomial
\begin{equation}\label{poly}
\gamma_{ij}\,x^ix^j=0
\end{equation}
together with the polynomial
\begin{equation}\label{conic}
\left(x^1\right)^2+\left(x^2\right)^2+\left(x^3\right)^2=0
\end{equation}
which is invariant under similarity transformations.

\section{A Diagram for the Vector-free Case}
The construction of a plane diagram for the vector-free case can be easily described using the concepts and terms of projective geometry.
However, since physicists tend not to be conversant with them
we use $3$-dimensional vector algebra remarking that the
points of the projective plane are nothing else but the rays originating from the origin of a $3$-dimensional Euclidean space. A ray is defined here as all positive and negative, but not zero, multiples of a non-zero vector.
Topologically the projective
plane is identical with the unit sphere on which opposite points have been identified.

A conic, like the one defined by \eqref{conic}, cannot  be mapped as a set of rays since it allows no real solutions. We interpret a symmetric matrix $a_{ij}$ with non-vanishing determinant up to a factor
\begin{equation}
\text{det}\left\| a_{ij}\right\|\ne0\,,\qquad a_{ij}=a_{ji}\,,
\end{equation}
as giving rise to the bilinear form
\begin{equation}
a_{ij}x^iy^j=0\,.
\end{equation}
To each ray $x^i$ belongs a covector
\begin{equation}
b_j=a_{ij}x^i
\end{equation}
up to a non-vanishing factor. We have then
\begin{equation}
b_j y^j=0 ,
\end{equation}
the equation of a plane of rays through the origin for the fixed ray $x^i$.

We apply this construction to the equation \eqref{conic} with
\begin{equation}
a_{ij}=\delta_{ij}
\end{equation}
and have the result that the  plane is orthogonal to the ray. Thus the ray
\begin{equation}
x^1=1\,,\quad x^2=0\,,\quad x^3=0
\end{equation}
gives rise to the plane
\begin{equation}
y^1=0
\end{equation}
which is the $x^2$--$x^3$ plane. In this way we obtain from the $3$ rays along the coordinate axes the $3$ coordinate planes orthogonal to them. Finally, we take a fourth ray not in the coordinate planes. We choose
\begin{equation}\label{rays}
x^1=1\,,\quad x^2=1\,,\quad x^3=1\,.
\end{equation}
It gives rise to a plane through the origin with equation
\begin{equation}
y^1+y^2+y^3=0\,.
\end{equation}
We turn now to our quadratic polynomial \eqref{poly} which characterizes the Ricci coefficients. Since \eqref{conic} is definite we can assume it to be in diagonal form
\begin{equation}
\gamma_1 \left(x^1\right)^2+\gamma_2 \left(x^2\right)^2+\gamma_3 \left(x^3\right)^2=0
\end{equation}
where $\gamma_j$ are real and determined up to a common, non-vanishing factor. We exclude the case that all $3$ $\gamma_j$ vanish. The ray\begin{equation}\label{rayx}
x^j=\gamma_j
\end{equation}
thus gives rise to the plane
\begin{equation}\label{plane}
\left(\gamma_1\right)^2 y^1+\left(\gamma_2\right)^2 y^2+\left(\gamma_3\right)^2 y^3=0\,.
\end{equation}
The intersection of this plane with the coordinate planes gives rise to the $3$ rays
\begin{equation}\label{polarity}
\left(\,0\,,\left(\gamma_3\right)^2,-\left(\gamma_2\right)^2\right),\; \left(\left(\gamma_3\right)^2,\,0\,,-\left(\gamma_1\right)^2\right),\; \left(\left(\gamma_2\right)^2,-\left(\gamma_1\right)^2,0\,\right).
\end{equation}
Since each ray \eqref{rayx} determines uniquely the plane \eqref{plane} the classification of the Ricci coefficients is that of the rays alone. Each ray can be identified with a pair of opposite points on the unit sphere, equivalent to a point on the real
 projective plane.

The map from the Ricci coefficients to the Ricci tensor given by equations \eqref{Gamma} and \eqref{rho} is not a map from rays into rays since the Ricci tensor remains invariant under space-reflection. To turn our map into an invertible one from the projective plane
$\mathbb{P}_X$ of the Ricci coefficient types to the Ricci tensor types we construct a real analog of a Riemann surface for the Ricci tensor types by extending the Ricci tensor into the Ricci tensor volume $R_{jk}/ \sqrt{\text{det}\left\| R_{
jk}\right\|}$. This results now in the ray map
\begin{equation}
y^j=\frac{1}{x^j}
\end{equation}
from the projective plane $\mathbb{P}_X$ into the projective plane $\mathbb{P}_Y$.

As long as all $x^j$ are different from zero the map is one-to-one and invertible. It creates a correspondence between rays not lying in the coordinate planes. This correspondence, well known to algebraic geometers as a Cremona transformation,
can be extended into a birational correspondence
from $\mathbb{P}_X$ to $\mathbb{P}_Y$ \cite{Mu}.
See Figure 2.

\begin{figure}[htb]\label{cremona}
\begin{center}
\includegraphics[width=0.90\textwidth]{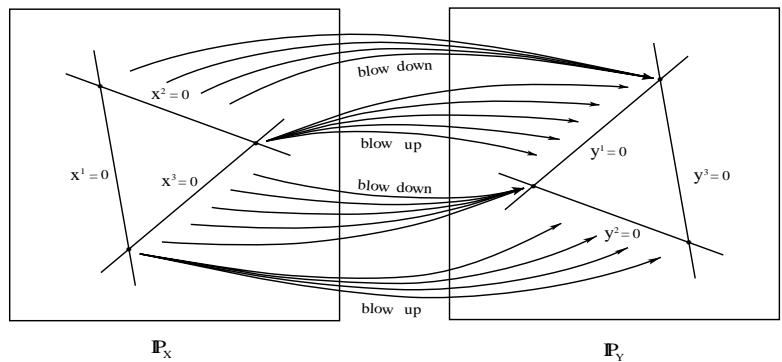}
\caption{The Cremona transformation is a special case of a birational map from $\mathbb{P}_X\backslash\{x^1x^2x^3\!=\!0\}$ to $\mathbb{P}_Y\backslash\{y^1y^2y^3\!=\!0\}$.
The graph shows the transformation of the coordinate planes.
The rays $x^i=x^j=0$ which are intersections of the coordinate planes
in $\mathbb{P}_X$ blow up to the
coordinate planes $y^k=0$ in $\mathbb{P}_Y$ with $i\ne j\ne k$ while the coordinate planes $x^i=0$ in $\mathbb{P}_X$ blow down to the intersections of coordinate planes $y^j=y
^k=0$ in $\mathbb{P}_Y$.}
\end{center}
\end{figure}

The rays $x^j$ and $y^k$ are subject to the equations
\begin{equation}
x^1y^1=x^2y^2=x^3y^3\,.
\end{equation}
If $x^1x^2x^3\ne0$ then at least one of the $y^k$ is different from zero and thus all $y^k$ are and we have
\begin{equation}
y^j=\frac{1}{x^j}\,.
\end{equation}
If $x^1=0$, $x^2x^3\ne0$ it follows that
\begin{equation}
y^2=y^3=0\,.
\end{equation}
This shows that a coordinate plane without its axes corresponds to the coordinate axis orthogonal to the plane.

Finally, if $x^1=x^2=0$ we have
\begin{equation}
y^3=0\,.
\end{equation}
To each coordinate axis corresponds the plane orthogonal to it.

Figure $3$. in the next section will show the correspondence.

\section{A Plane Diagram}
To obtain a diagram of the projective plane we use a gnomic projection of the rays through the origin to the plane
\begin{equation}
x^1+x^2+x^3=1\,.
\end{equation}
This means that we normalize the $\gamma_j$ by introducing the $a_k$
\begin{equation}
a_k=\frac{\gamma_k}{ \sum_j \gamma_j }\,,\qquad \sum_{k} a_k=1\,.
\end{equation}
We interpret then the $a_k$ as triangular coordinates. The planes $x^j=0$ become now the sides of an equilateral triangle of height $\sqrt{\frac{3}{2}}$ and the $a_k\sqrt{\frac{3}{2}}$ measure the distances from the side $x^k=0$.
The coordinates are counted positive towards the interior of the triangle.
See Figure $3$.

\begin{figure}[htb]\label{viviani}
\begin{center}
\includegraphics[width=0.60\textwidth]{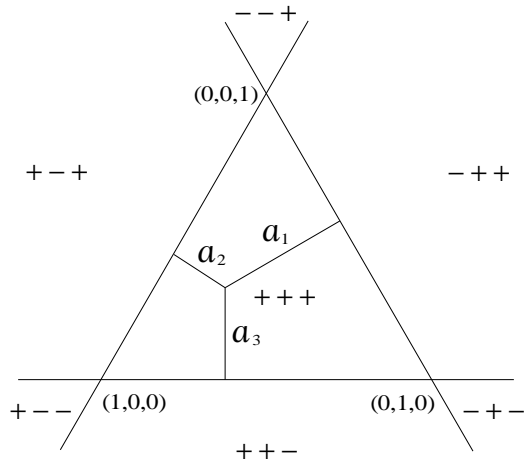}
\caption{The graph shows a gnomic projection of the rays through the origin to the plane $x^1+x^2+x^3=1$.}
\end{center}
\end{figure}

A simple construction for the lines \eqref{polarity} is given by Coxeter \cite{hsmC} and will not be repeated here. The triangular coordinates are preferred here since they reflect the symmetry of the classification under permutation of the $3$ eigenvalues
$\gamma_j$. However, we have to admit also values of the $a_k$ on the line at infinity since $\sum \gamma_j$ is allowed to vanish.

From \eqref{Riccistructure} it is clear that a classification of the Ricci coefficients under similarity implies a corresponding one for the structure constants of the group.

Finally, we remark that the constant Ricci coefficients are examples of what we have called ``constant gravitational fields'' which we have treated in a book to be published \cite{SS}.


\end{document}